\documentclass[12pt]{article}
\usepackage{epsfig}
\usepackage{amsmath}
\makeatletter
\def\gsim{\compoundrel>\over\sim}
\def\lsim{\compoundrel<\over\sim}
\def\compoundrel#1\over#2{\mathpalette\compoundreL{{#1}\over{#2}}}
\def\compoundreL#1#2{\compoundREL#1#2}
\def\compoundREL#1#2\over#3{\mathrel{\vcenter{\hbox{$\buildrel{#1#2}\over{#1#3}$}}}}
\begin{document}
\setcounter{page}{1}
\setlength{\parindent}{1.0em}
\begin{flushright}
GUTPA/03/04/01
\end{flushright}
\renewcommand{\thefootnote}{\fnsymbol{footnote}}
\begin{center}{\LARGE{{\bf Supersymmetric Grand Unification with
a Fourth Generation?}}}\end{center}
\begin{center}{\large{J. E. Dubicki}~\footnote[2]{E-mail:
j.dubicki@physics.gla.ac.uk} \; \large{and C. D.
Froggatt}~\footnote[3]{E-mail: c.froggatt@physics.gla.ac.uk}\\}
\end{center}
\renewcommand{\thefootnote}{\arabic{footnote}}
\begin{center}{{\it Department of Physics and Astronomy}\\{\it
University of Glasgow, Glasgow G12 8QQ, Scotland}}\end{center}
\setcounter{footnote}{0}
\begin{abstract}
The possibility of incorporating a chiral fourth generation into a
SUSY-GUT is investigated. Precision fits to electroweak
observables require us to introduce light supersymmetric
particles, with masses less than $M_Z$. These particles might also
provide decay channels for the fourth generation quarks of mass
$\sim$ 100 GeV. We also require $\tan\beta$ to lie in the range
1.50 $\lsim \tan\beta \lsim$ 1.75 and obtain an upper limit on the
lightest Higgs boson mass in the MSSM4 of 152 GeV.
\end{abstract}
\section{Introduction}
\label{sec1} The Standard Model (SM) is, of course,
phenomenologically very successful, while its supersymmetric
extension stabilizes the gauge hierarchy problem and allows a
grand unification of the SM interactions. However, there is no
explanation of why there should be just three generations of
quarks and leptons or their hierarchy of masses. In this paper we
investigate the possibility of consistently incorporating a fourth
generation into a supersymmetric grand unified theory.\par We
shall assume a structure akin to that of the minimal
supersymmetric standard model (MSSM3), adding a complete chiral
fourth generation and its associated supersymmetric partners (the
so-called MSSM4). We note that at one-loop level the validity of
gauge coupling unification is independent of the number of
generations, and so this requirement does not discriminate between
three and four generation models. However, the requirement that
all the Yukawa couplings remain perturbative at energies up to the
GUT scale places restrictive upper bounds on the fourth generation
$T$, $B$, $E$ and $N$ masses and also constrains the
supersymmetric parameter $\tan \beta =
\frac{\upsilon_U}{\upsilon_D}$ ($\upsilon_U$ and $\upsilon_D$
being the vacuum expectation values of the two Higgs doublets
$H_U$ and $H_D$ that are present in the MSSM). In
section~\ref{sec2} we investigate the masses and decay channels of
the fourth generation particles that are consistent both with
perturbative gauge coupling unification and the latest
experimental bounds from direct searches. In section~\ref{sec3} we
comment on the issues regarding fits to precision data in the
MSSM4 before discussing the lightest Higgs boson mass in
section~\ref{sec4}.\par
\section{Experimental Constraints and a Fourth Generation}
\label{sec2} All fourth generation models must adhere to certain
experimental constraints, the first of which stems from precise
measurements of the decay characteristics of the $Z$-boson
performed at LEP. This has set a lower bound of
$M_F\ge\frac{M_Z}{2}$ on any non-SM particles that couple to the
$Z$-boson. Ignoring the unnatural hierarchy emerging within the
neutrino sector, we assume a Dirac mass $M_N
\sim\left(\frac{M_Z}{2}\right)$ for the heavy neutral lepton.
\par We begin our discussion by considering the
leptonic sector where we assume $M_E > M_N$. Under the assumption
that the mixing between the fourth generation leptonic sector and
the first three generations is negligible, the decay $E
\rightarrow N W^*$ will be dominant. Current experimental limits
searching for $E \rightarrow N W^*$ from $e^+e^- \rightarrow
E^+E^-$ production have been performed by the OPAL and L3
collaborations up to the kinematic limit $M_E \sim$ 100
GeV~\cite{lepton}. However, it turns out that to be consistent
with perturbative unification we require $M_E \sim \frac{M_Z}{2}$,
so that the fourth generation charged lepton Yukawa coupling $Y_E$
doesn't become non-perturbative below the GUT scale. Therefore, in
order to evade experimental bounds, the mass difference $\Delta
M_L$ must be less than $\sim$ 5 GeV, which is in the region where
the trigger efficiencies are significantly lowered and events are
dominated by the two photon background~\cite{lepton}. We also note
that with $\Delta M_L$ the order of a few GeV the decay lifetime
$\tau \left( E \rightarrow N W^{\ast} \right)$ is too short for the 
heavy $E$ lepton to leave a charged track in the electromagnetic
calorimeter. Regarding
the heavy neutrino, OPAL and L3 have set the bound $M_N >$ 70 - 90
GeV based on the search for $N \rightarrow l W^*$ ($l=e, \mu$ or
$\tau$) provided the mixing matrix elements satisfy
$V_{Ne,\mu,\tau} > 10^{-6}$~\cite{lepton}. However, if we assume
that this mixing angle is negligible ($V_{Ne,\mu,\tau} <
10^{-6}$), then the neutrino is stable enough to leave the
detector. In this case, the only relevant bound comes from the
$Z^0$ decay width noted earlier. Based on the above discussion, in
the rest of this paper we shall take:
\begin{equation}
    M_E \simeq 50 \; \mbox{GeV} \quad ; \quad M_N \simeq 50 \;
    \mbox{GeV}
\end{equation}
as representative masses of the fourth generation leptonic
particles.\par Direct searches for the fourth generation quarks is
an ongoing process at the Fermilab Tevatron. Here we focus on the
experimental restrictions for $M_t > M_T > M_B$, so that the
charged current (CC) decays $B \rightarrow t W^-$ and $B
\rightarrow T W^-$ are kinematically forbidden. The leading CC
decay mode will then be $B \rightarrow c W^-$ which is doubly
Cabibbo suppressed by the mixing matrix factor $V_{cB}$. In this
situation, loop induced flavour changing neutral current (FCNC)
decays can dominate provided~\cite{hou}:
\begin{equation}
    \frac{ \left| V_{cB} \right| }{ \left| V_{tB} \right| } \lsim
    {\cal O} \left( 10^{-2} - 10^{-3} \right)
\end{equation}
Several experiments have searched explicitly for $B$ quarks
decaying via FCNCs. The $D \emptyset$ collaboration~\cite{do} has
excluded the range $\frac{M_{Z^0}}{2} < M_B < M_{Z^0} + M_b$ by a
null search for both $B \rightarrow b \gamma$ and $B \rightarrow b
g$. For masses $M_t,M_T > M_B > M_{Z^0} + M_b$, the decay $B
\rightarrow b Z^0$ is expected to dominate, except for $B
\rightarrow b h^0$, if $M_B > M_{h^0} + M_b$. The CDF
collaboration~\cite{cdf} has performed a general search for
long-lived particles that decay into a $Z^0$ gauge boson. This
will encompass the FCNC decay of a fourth generation $B$ quark
decaying via $B \rightarrow b Z^0$, if the mixing matrix factor
$V_{tB} \simeq V_{Tb}$ is small enough to result in a long
lifetime. By looking for $Z^0 \rightarrow e^+e^-$ with a displaced
vertex they are able to exclude a $B$ quark mass up to $M_B=$ 148
GeV for $c \tau_B =$ 1 cm, where $\tau_B$ is the proper decay time
of the $B$ quark, and a branching ratio of $Br \left( B
\rightarrow b Z^0 \right) =$ 100 \%. However, this limit
diminishes to $M_B \sim$ 96 GeV if $c \tau_B >$ 22 cm. To date,
this remains the only lower mass bound on quasi-stable $B$ quarks
as emphasized by Frampton {\it et al.}~\cite{frampton}.\par As
regards the $T$ quark, for $M_T \gsim$ 100 GeV, there are two
competing decay modes; $T \rightarrow B W$ and $T \rightarrow b
W$. The $BW$ decay will certainly dominate over the $bW$ decay
when the former is real, since the $bW$ channel is suppressed by
the mixing matrix factor $V_{Tb}$. However, when $M_T-M_B<M_W$,
the two body $bW$ decay could be competitive with the three body
$BW^{\ast}$ decay. For large enough $V_{Tb}$ the $T \rightarrow
bW$ decay will be dominant. However, for $M_T \lsim M_t$, such a
dominance would lead to a large excess of $bW$ events relative to
those already present from $t \overline{t}$ production and decay,
and so this implies that $\frac{V_{Tb}}{V_{TB}}$ must be small
enough for the $T \rightarrow BW^{\ast}$ decay to dominate. The
detection of the $T$ quark would then depend on the decay
properties of the $B$ quark and the mass difference $M_T - M_B$.
We refer to~\cite{gunion} and~\cite{hung} for a discussion of this
scenario, in particular Gunion {\it et al.}~\cite{gunion} finds
the only way that $T \overline{T}$ events can evade being included
in the CDF and $D \emptyset$ data sample is if $M_T - M_B$ is
sufficiently small so that the $W^{\ast}$ in $T \rightarrow
BW^{\ast}$ is virtual and the jets and leptons from the two
$W^{\ast}$'s are soft. A further analysis on updated data is
needed that takes into account the quasi-stable nature of the $B$
quark.\par
\begin{figure}[p]
\centerline{\epsfig{file=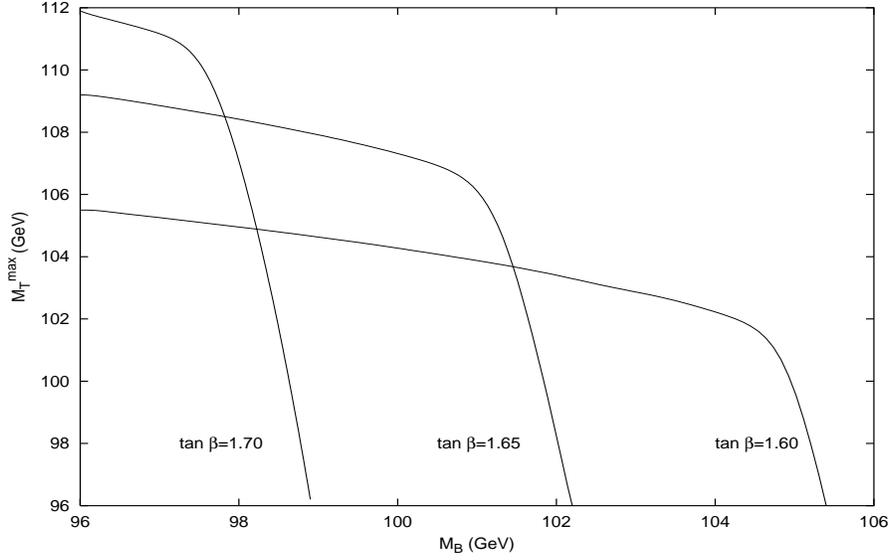,height=7.5cm,width=12cm}}
\caption{\small{Plot of $M_T^{max}$ versus $M_B$ for $\tan \beta =
$ 1.60, 1.65 and 1.70, as obtained from the requirement of
perturbative gauge coupling unification. The third generation $t$
quark mass is taken as $M_t =$ 175 GeV. The fourth generation
leptonic masses are taken as $M_E = M_N =$ 50 GeV.}}
\label{mtmaxuni}
\end{figure}

 To obtain the allowed masses of the fourth generation
quarks that are consistent with perturbative gauge coupling
unification we perform a renormalization group study of the MSSM4.
Specifically, this enables us to place upper limits on the masses
of the $T$ and $B$ quarks by ensuring their Yukawa couplings $Y_T$
and $Y_B$ run perturbatively to the GUT scale $M_{GUT}$:
\begin{equation}
    Y^2_{T,B} \left( \mu \right) \leq 4 \pi \quad \mbox{for} \quad
    M_{Z^0} \leq \mu \leq M_{GUT}
\end{equation}
where $M_{GUT}$ is defined to be the scale where $\alpha_1 ( \mu )
= \alpha_2 ( \mu )$. In our analysis we have neglected all the
Yukawa couplings from the first three generations, except that of
the third generation $t$ quark whose mass we take to be $M_t =$
175 GeV. As is typical with four generation models, we also
require small values of $\tan \beta$ so as to avoid $Y_B \left(
M_{Z^0} \right) \geq {\cal O} \left( \sqrt{4 \pi} \right)$.
Further details of the procedure for running the renormalization
group equations in the MSSM4 can be found in~\cite{josefd}. In
Figure~\ref{mtmaxuni} we have plotted the maximum $T$ quark mass
($M_T^{max}$) versus $M_B$, given that perturbative unification
must occur.
\begin{figure}[p]
\centerline{\epsfig{file=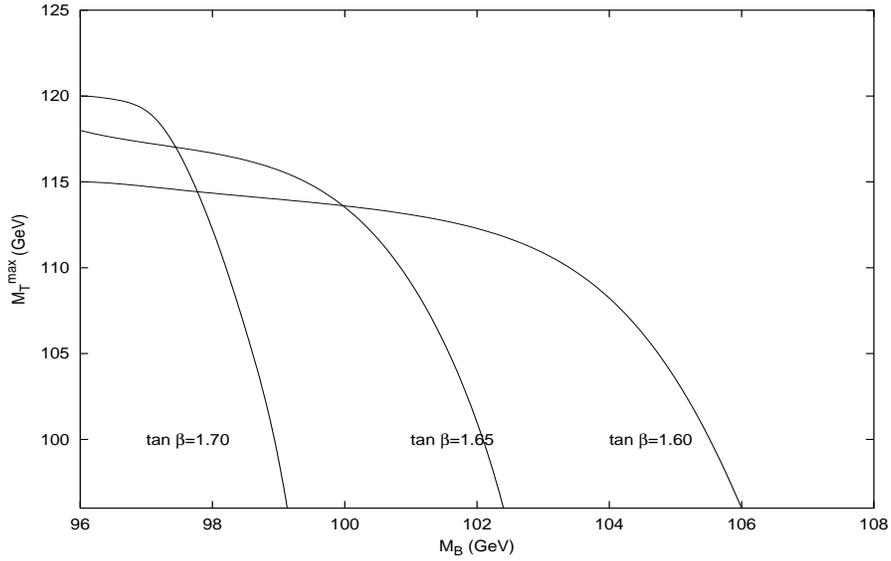,height=7.5cm,width=12cm}}
\caption{\small{Same as Figure~\ref{mtmaxuni}, but with $M_t =$
170 GeV.}} \label{mtmaxsecondvaluemt}
\end{figure}
We plot for $\tan \beta =$ 1.60, 1.65 and 1.70, and we fix $M_t =$
175 GeV. We have taken $M_B \geq$ 96 GeV, which corresponds to the
absolute experimental lower bound on a quasi-stable $B$ quark as
discussed earlier. We can see from Figure~\ref{mtmaxuni} that
higher $\tan \beta$ values allow for a larger $M_T^{max}$, though
at the expense of restricting $M_B$, and as such tends to favour
the hierarchy:
\begin{equation}
    M_T > M_B
\label{eqhierarchy}
\end{equation}
This mass hierarchy is necessary if we consider the $T$ and $B$
quarks to have the standard model decay channels $T \rightarrow B
W^{\ast}$ and $B \rightarrow b Z^0$, as discussed earlier in this
section~\footnote{If $M_T < M_B$ then we would expect $T
\rightarrow b W$ to be the dominant decay channel of the $T$
quark, which is excluded experimentally based on searches for the
third generation top quark.}. As we decrease $\tan \beta$ then the
hierarchy in Eq.~\ref{eqhierarchy} becomes harder to maintain.
Overall, we find that $\tan \beta$ can take on the values 1.50 $<
\tan \beta <$ 1.75 without the Yukawa couplings becoming
non-perturbative below the GUT scale. In
Figure~\ref{mtmaxsecondvaluemt}, we also plot $M_T^{max}$ versus
$M_B$, but this time using $M_t =$ 170 GeV. We can see that this
value of $M_t$ increases $M_T^{max}$ by a few GeV for a given
$M_B$ and $\tan \beta$. For $M_t =$ 180 GeV the allowed ranges of
$M_T$ and $M_B$ are very constrained, in fact we only find
solutions for $\tan \beta \geq$ 1.66 and $M_T , M_B <$ 100 GeV.
\par Looking at Figures~\ref{mtmaxuni}
and~\ref{mtmaxsecondvaluemt} it is clear that we could arrange
$M_T$ and $M_B$ so that the standard model decay channels:
\begin{equation}
    T \rightarrow B W^{\ast} \quad ; \quad B \rightarrow b Z^0
\end{equation}
are kinematically accessible, and dominate on the basis that the
mixing angles $V_{Bc}$ and $V_{Tb}$ are suppressed. An updated
analysis based on RUNII data at Fermilab is essential to exclude this
scenario. However, we also note that there are other decay
channels for the $T$ and $B$ quarks that become possible in the
MSSM4. For instance, we can consider the possibility of light
supersymmetric particles providing decay channels for the $T$ and
$B$ quarks. In this situation one can constrain the masses of the
light ({\it i.e.} $< M_{Z^0}$) neutralino (${\tilde{\chi}}^0_1$)
and chargino (${\tilde{\chi}}^{\pm}_1$) pair which, as we will see
in section~\ref{sec3}, is already required by fits to precision
data, in order to allow the following two body decays:
\begin{equation}
    T \rightarrow {\tilde{B}}_1 {\tilde{\chi}}^+_1 \quad ; \quad B
    \rightarrow {\tilde{B}}_1 {\tilde{\chi}}^0_1
    \label{susydecay}
\end{equation}
where ${\tilde{B}}_1$ is the lightest mass eigenvalue in the
fourth generation sbottom sector. It has already been shown by
Gunion {\it et. al.}~\cite{gunion} that ${\tilde{B}}_1$ is
typically the lightest squark in the MSSM4. Ensuring both of the
decay channels in Eq.~\ref{susydecay} are kinematically
accessible, combined with the constraints from perturbative
unification and precision data fits, places severe restrictions on
the allowed spectrum. We discuss this in more detail in
section~\ref{sec3}.
\section{Precision Measurements and a Fourth Generation}
\label{sec3} It is difficult to provide bounds from precision data
without a fully consistent study taking into account exact
particle masses, any light supersymmetric particle spectra present
and mixings between different flavours. However, Maltoni {\it et.
al.}~\cite{maltoni} has pointed out that a highly degenerate
neutralino and chargino pair can provide the necessary
contributions to the precision parameters that are needed to
cancel that of the fourth generation, whilst at the same time
being consistent with LEP bounds~\cite{degneut}. Specifically they
require:
\begin{equation}
    M_{\pi^+} \lsim \Delta M_{\tilde{\chi}} \lsim 3 \; \mbox{GeV}
    \quad \mbox{and} \quad M_{\tilde{\chi}} \sim 60 \; \mbox{GeV}
\end{equation}
where we define the notation $\Delta M_{\tilde{\chi}} =
M_{{\tilde{\chi}}^{\pm}_1} - M_{{\tilde{\chi}}^0_1}$ and
$M_{\tilde{\chi}} = M_{{\tilde{\chi}}^{\pm}_1} \gsim
M_{{\tilde{\chi}}_1^0}$. Looking at their results, we see that the
magnitude of the contribution to the fitted parameters from this
sector is highly dependent on $M_{\tilde{\chi}}$. Deviations from
$M_{\tilde{\chi}}$ larger than + 30 GeV or - 5 GeV are ruled out
at the $ 2 \sigma$ level. We now discuss the influence of these
precision data fits on the scenario where the $T$ and $B$ quarks
decay into supersymmetric particles.\par If allowed, the $B \rightarrow
{\tilde{B}}_1 {\tilde{\chi}}^0_1$ decay in Eq.~\ref{susydecay}
will certainly dominate over the one-loop FCNC decay $B
\rightarrow b Z^0$ and two-generation decays $B \rightarrow c W^-$
that traditional searches have looked for. Typical (pole) masses
required for this scenario to be viable are:
\begin{eqnarray}
    M_B \left( \simeq M_T \right) & \gsim & 100 \; \mbox{GeV} \nonumber \\
    M_{{\tilde{B}}_1}& \lsim & 50 \; \mbox{GeV} \label{soltbdecay} \\
    55 \; \mbox{GeV} & \leq & M_{\tilde{\chi}} \leq M_B -
    M_{{\tilde{B}}_1} \nonumber
\end{eqnarray}
We assume that $\theta_{\tilde{B}} \simeq$ 1.17 rad, where
$\theta_{\tilde{B}}$ is the mixing angle that relates the
${\tilde{B}}_L$ and ${\tilde{B}}_R$ weak eigenstates to the mass
eigenstates ${\tilde{B}}_1$ and ${\tilde{B}}_2$ through the
matrix:
\begin{equation}
    \left( \begin{array}{c} {\tilde{B}}_1 \\ {\tilde{B}}_2
    \end{array} \right) = \left( \begin{array}{cc} \cos
    \theta_{\tilde{B}} & \sin \theta_{\tilde{B}} \\ - \sin
    \theta_{\tilde{B}} & \cos \theta_{\tilde{B}} \end{array}
    \right) \left( \begin{array}{c} {\tilde{B}}_L \\ {\tilde{B}}_R
    \end{array} \right)
\end{equation}
This value of $\theta_{\tilde{B}}$ is chosen so that
${\tilde{B}}_1$ decouples from the $Z^0$-boson, thereby
maintaining consistency with the measured total $Z^0$ width if
$M_{{\tilde{B}}_1} < \frac{M_{Z^0}}{2}$. It turns out that
$\theta_{\tilde{B}} \simeq$ 1.17 rad is also required in order to
evade the experimental direct searches for the light
${\tilde{B}}_1$ squark. We discuss this in more detail later in
this section.\par The decay rates for $T \rightarrow {\tilde{B}}_1
{\tilde{\chi}}^+_1$ and $B \rightarrow {\tilde{B}}_1
{\tilde{\chi}}^0_1$ depend crucially on the amount of
${\tilde{B}}_L - {\tilde{B}}_R$ mixing. It is especially important
to consider the effect of fixing $\theta_{\tilde{B}} \simeq$ 1.17
rad on the $T \rightarrow {\tilde{B}}_1 {\tilde{\chi}}^+_1$ decay.
The Lagrangian for the ${\tilde{B}}_1$-$T$-${\tilde{\chi}}^{+}_1$
interaction is given by~\cite{haber}:
\begin{eqnarray}
    {\cal L}_{T {\tilde{B}}_1 {\tilde{\chi}^+_1}} & = & g
    \overline{T} P_R \left( - U_{11} \cos \theta_{\tilde{B}} +
    U_{12} Y_B \sin \theta_{\tilde{B}} \right) {\tilde{\chi}}_1^+
    {\tilde{B}}_1 \nonumber \\
    & & + g \overline{T} P_L \left( Y_T V_{12} \cos
    \theta_{\tilde{B}} \right) {\tilde{\chi}}_1^+
    {\tilde{B}}_1 + h.c.
    \label{intqsqucharg}
\end{eqnarray}
where $P_{L,R} = \frac{1}{2} \left( 1 \mp \gamma_5 \right)$ and
$U_{ij}$ and $V_{ij}$ ($i,j = 1,2$) are the $2 \times 2$ unitary
matrices diagonalizing the charged gaugino-higgsino
matrix~\footnote{$\mu_H$ is the bilinear term that couples the two
Higgs doublets in the superpotential, whilst $M_2$ is the $SU(2)$
gaugino soft mass that appears in the supersymmetric breaking
Lagrangian.}:
\begin{equation}
    \textbf{U}^{\ast} \left( \begin{array}{cc} M_2 & \sqrt{2} \sin
    \beta M_{W} \\ \sqrt{2} \cos \beta M_{W} & \mu_H \end{array}
    \right) \textbf{V}^{-1} = \left( \begin{array}{cc}
    M_{{\tilde{\chi}}_1^{\pm}} & 0 \\ 0 &
    M_{{\tilde{\chi}}_2^{\pm}} \end{array} \right)
\end{equation}
We take $\mu_H$ and $M_2$ to be in the parameter regions where we
expect a light degenerate chargino-neutralino pair, required in
the MSSM4 by fits to precision data~\cite{maltoni}. If we take
$M_2 \ll \left| \mu_H \right|$ then the ${\tilde{\chi}}^0_1$ and
${\tilde{\chi}}^{\pm}_1$ are both gaugino-like, with
$M_{{\tilde{\chi}}^{\pm}_1} \simeq M_2$. In this case the mass of
the lightest neutralino is given by $M_{{\tilde{\chi}}^0_1} \simeq
\min \left( M_1 , M_2 \right)$, where $M_1$ is the $U(1)$ gaugino
soft mass. Therefore, small mass splittings only occur if $M_2 <
M_1$. The mixing matrix elements in the chargino sector for the
gaugino-like case are given by:
\begin{equation}
    V_{11} \sim U_{11} \sim 1 \quad V_{12} \sim U_{12} \sim 0
\end{equation}
On the other hand, if $\left| \mu_H \right| \ll M_2$ then
${\tilde{\chi}}^0_1$ and ${\tilde{\chi}}^{\pm}_1$ are both
higgsino-like with degenerate masses $M_{{\tilde{\chi}}^{\pm}_1}
\simeq M_{{\tilde{\chi}}^0_1} \simeq \left| \mu_H \right|$. In
this case, the mixing matrix elements in the chargino sector obey:
\begin{equation}
    V_{11} \sim U_{11} \sim 0 \quad V_{12} \sim sgn \left( \mu_H
    \right) \quad U_{12} \sim 1
\end{equation}
We can see from Eq.~\ref{intqsqucharg} that the decay rate for $T
\rightarrow {\tilde{B}}_1 {\tilde{\chi}}_1^+$ will be suppressed
if the light chargino ${\tilde{\chi}}_1^+$ is gaugino-like, since
$V_{12} \sim U_{12} \sim 0$ and $\cos \theta_{\tilde{B}}$ is fixed
at 0.39. Therefore, in order to ensure that $T \rightarrow
{\tilde{B}}_1 {\tilde{\chi}}^+_1$ dominates over $T \rightarrow b
W^+$, without appealing to a suppression of the CKM mixing matrix
element $V_{Tb}$, we would expect the light chargino
${\tilde{\chi}}_1^+$ to be higgsino-like.\par

As with most phenomenological studies of supersymmetric models
with $R$-parity, we must ensure that the lightest supersymmetric
particle (LSP) is neutral for cosmological reasons. This is
usually taken to be the neutralino ${\tilde{\chi}}^0_1$ in the
MSSM3. The masses in Eq.~\ref{soltbdecay} contradict this
requirement. One possible solution presents itself, however, if we
assume that the LSP is in fact the fourth generation sneutrino
${\tilde{N}}_1$ which becomes stable due to
$R$-parity~\footnote{Previous studies of the MSSM4 have shown that
for some of the parameter space it is natural to assume that
${\tilde{N}}_1$ is the LSP~\cite{gunion}.}. Since
$M_{{\tilde{N}}_1} < \frac{M_{Z^0}}{2}$ we must arrange the mixing
angle $\theta_{\tilde{N}}$ in the fourth generation sneutrino mass
matrix such that ${\tilde{N}}_1$ decouples from the $Z^0$-boson. A
right-handed LSP ${\tilde{N}}_1$ does indeed not couple to the
$Z^0$-boson. The light fourth generation bottom squark will then
decay via the semi-leptonic channel ${\tilde{B}}_1 \rightarrow c l
{\tilde{N}}_1^{\ast}$, where $l=e, \mu , \tau$. Such a decay
involves the factor $ \left| V_{Bc} V_{N l} \right| $, leading to
a long lifetime. Experiments looking for hadronizing sbottom
squarks currently only exclude the range 5 GeV $ \leq
M_{{\tilde{B}}_1} \leq $ 38 GeV, if the mixing angle
$\theta_{\tilde{B}} \simeq$ 1.17 rad such that ${\tilde{B}}_1$
decouples from the $Z^0$-boson~\cite{delphi1}. Searches for long
lived charged particles in $p \overline{p}$ collisions at CDF have
only excluded particles in the mass range 50 GeV - 270
GeV~\cite{cdf2}.
\section{The Higgs Sector of the MSSM4}
\label{sec4} At tree-level the lightest Higgs boson mass $M_{h^0}$
in the MSSM4 is bounded from above by the relation:

\begin{eqnarray}
    M_{h^0} & \leq & M \left| \cos 2 \beta \right| \leq
    M_{Z^0} \nonumber \\
    M & \equiv & \min \left( M_{Z^0} , M_{A^0} \right)
\end{eqnarray}

This has already been ruled out from experimental data taken at
LEP~\cite{raspereza}. It is therefore of vital importance to check
that the radiative corrections in the MSSM4 can provide a large
enough contribution to raise the mass of the lightest Higgs boson
to be above its experimental lower bound.\par To take the
radiative corrections into account in the MSSM4, we use the
one-loop effective potential, given by:
\begin{equation}
    V_{eff} = V_0 + V_1 + \ldots
\end{equation}
where $V_0$ is the tree-level potential, $V_1$ contains the
one-loop contributions, and the ellipses represent the higher-loop
corrections which we shall ignore. The one-loop radiative
correction to the effective potential in the MSSM4 is given by:
\begin{equation}
    V_1 = \sum_k \frac{1}{64 \pi^2} {(-1)}^{2 J_k} (2 J_k + 1) C_k
    m_k^4 \left( \ln \frac{m_k^2}{Q^2} - \frac{3}{2} \right)
    \label{oneloopeff}
\end{equation}
where the sum is taken over all particles in the loop; $C_k=6(2)$
for coloured (uncoloured) fermions; $J_k$ are the spins and $m_k$
are the field dependent masses of the particles in the loops at
the renormalization scale $Q$. We choose to minimize the potential
at the scale $Q=\max(M_t,M_T)$.\par In this analysis we will
consider contributions to $V_1$ that arise from the $t$, $T$, $B$,
$E$ and $N$ fermions and their corresponding superpartners. Since
we are at low $\tan \beta$ we can ignore contributions to $V_1$
from the third generation (s)bottom (s)quark. To calculate the
lightest Higgs boson mass we make the approximations:
\begin{equation}
    M_{\tilde{l}} , M_{A^0} , M_{\tilde{q}} = {\cal O} \left( M_{S} \right)
    \quad ; \quad l=E,N \quad ; \quad q=t,T,B 
\end{equation}
where $M_S$ represents the scale of the soft supersymmetric breaking terms. Further
details of the procedure can be found in~\cite{josefd}.\par
\begin{figure}[t]
\centerline{\epsfig{file=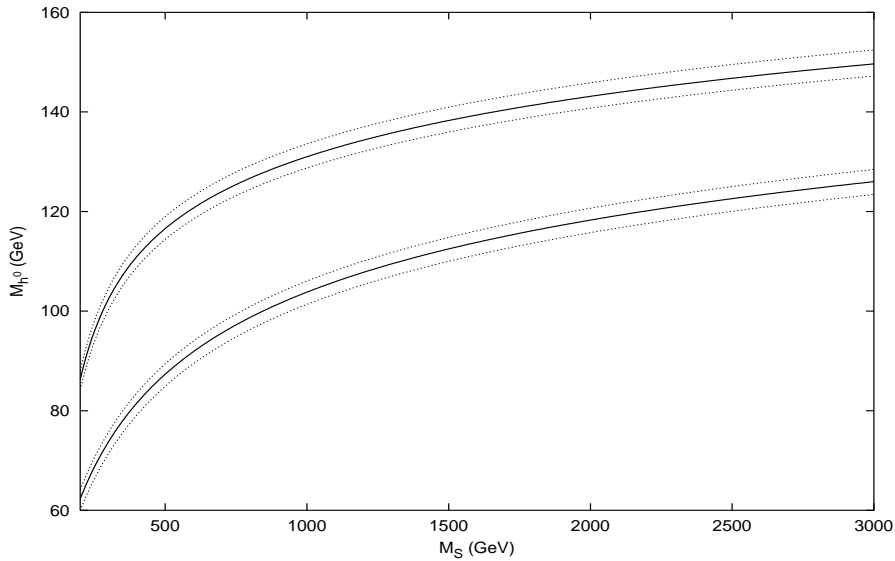,height=7.5cm,width=12cm}}
\caption{\small{Plot of the lightest Higgs boson mass $M_{h^0}$
versus $M_S$. The upper and lower solid curves correspond to $M_T
= M_B =$ 100 GeV ; $M_E = M_N =$ 50 GeV and $\tan \beta =$ 1.65
for the maximal-mixing and no-mixing cases, respectively. The
dotted curves correspond to the uncertainty in $M_{h^0}$ from
allowing $M_T$, $M_B$ and $\tan \beta$ to take on any value as
long as perturbative unification occurs.}} \label{plot7}
\end{figure}
In Figure~\ref{plot7} we plot the mass of the lightest Higgs boson
mass $M_{h^0}$ versus $M_S$ for $M_t =$ 175 GeV, and values of
$M_T$, $M_B$ and $\tan \beta$ that results in perturbative
unification. The bold curves correspond to taking $M_T = M_B =$
100 GeV, $M_E = M_N =$ 50 GeV and $\tan \beta =$ 1.65. We have
also determined the allowed $\left( M_T , M_B , \tan \beta
\right)$ parameter space that results in perturbative unification,
and from these sets of values we calculate the lightest Higgs
boson mass. For a given $M_S$, we retain the maximum and minimum
values of $M_{h^0}$ returned, which are represented by the dotted
curves. We have also plotted for the maximal-mixing and
minimal-mixing scenarios, as discussed in detail by Espinosa~\cite{espinosa}.
Overall, we can see that the maximum Higgs boson mass
$M_{h^0}^{upper}$ consistent with perturbative
unification has a value of $M_{h^0}^{upper} \simeq$ 152 GeV. This
is safely above the LEP lower bound.
\section{Conclusions}
\label{sec5} We have seen that is it possible to incorporate a
fourth generation into a supersymmetric GUT model, requiring the
existence of a light, degenerate neutralino and chargino pair in
order to provide the necessary cancellations in precision data
fits. The fourth generation masses are then tightly constrained
with typical values of $M_T \simeq M_B \simeq$ 100 GeV and $M_E
\simeq M_N \simeq$ 50 GeV. To retain perturbative consistency to
the unification scale we constrain $\tan \beta$ to lie in the
interval 1.50 $\lsim \tan \beta \lsim$ 1.75. In order to provide
decay channels to the fourth generation quarks, it might be that
the LSP is the sneutrino ${\tilde{N}}_1$ with a mass
$M_{{\tilde{N}}_1} \lsim$ 40 GeV. We would also expect a light
${\tilde{B}}_1$ with $M_{{\tilde{B}}_1} \simeq$ 40--50 GeV along
with a degenerate neutralino and chargino pair $M_{\tilde{\chi}}
\simeq$ 55--60 GeV. Such a light ${\tilde{B}}_1$ squark is chosen
to decouple from the $Z^0$-boson, but would be copiously produced
at the Fermilab Tevatron and should be searched for.
The upper limit on the lightest Higgs boson
mass is $M_{h^0}^{upper} \simeq$ 152 GeV. The supersymmetric
spectrum needed to satisfy all these constraints cannot be
obtained from MSUGRA scenarios with universal parameters at the
unification scale. We should also remark that a large degree of fine-tuning 
of parameters is involved in such a model.\par
\section*{Acknowledgements}
We would like to thank D. G. Sutherland for useful discussions. One of the 
authors (JED) was supported by a PPARC studentship.

\end{document}